\documentclass[ammsymb,prl,aps,twocolumn,superscriptaddress,showpacs]{revtex4}
\usepackage{graphicx}

\def \beq{\begin{equation}}
\def \eeq{\end{equation}}
\def \beqar{\begin{eqnarray}}
\def \eeqar{\end{eqnarray}}
\input{tcilatex}

\begin{document}

\title{Abrupt field-induced transition triggered by magnetocaloric effect in
phase-separated manganites}
\author{L. Ghivelder}
\affiliation{Instituto de F\'{\i}sica, Universidade Federal do Rio de Janeiro, C.P.
68528, Rio de Janeiro, RJ 21941-972, Brazil}
\author{R.S. Freitas}
\affiliation{Instituto de F\'{\i}sica, Universidade Federal Fluminense, C.P. 68528, Niter%
\'{o}i, RJ 21945-970, Brazil}
\author{M.G. das Virgens}
\affiliation{Instituto de F\'{\i}sica, Universidade Federal do Rio de Janeiro, C.P.
68528, Rio de Janeiro, RJ 21941-972, Brazil}
\affiliation{Instituto de F\'{\i}sica, Universidade Federal Fluminense, C.P. 68528, Niter%
\'{o}i, RJ 21945-970, Brazil}
\author{M.A. Continentino}
\affiliation{Instituto de F\'{\i}sica, Universidade Federal Fluminense, C.P. 68528, Niter%
\'{o}i, RJ 21945-970, Brazil}
\author{H. Martinho}
\affiliation{Instituto de F\'{\i}sica Gleb Wataghin, UNICAMP, Campinas, SP 13083-970,
Brazil}
\author{L. Granja}
\author{M. Quintero}
\affiliation{Departamento de F\'{\i}sica, Comisi\'{o}n Nacional de Energ\'{\i}a At\'{o}%
mica, Av. Gral Paz 1499 (1650) San Mart\'{\i}n, Buenos Aires, Argentina}
\author{G. Leyva}
\affiliation{Departamento de F\'{\i}sica, Comisi\'{o}n Nacional de Energ\'{\i}a At\'{o}%
mica, Av. Gral Paz 1499 (1650) San Mart\'{\i}n, Buenos Aires, Argentina}
\affiliation{Escuela de Ciencia y Tecnolog\'{\i}a, UNSAM, Alem 3901, San Mart\'{\i}n,
Buenos Aires, Argentina}
\author{P. Levy}
\affiliation{Departamento de F\'{\i}sica, Comisi\'{o}n Nacional de Energ\'{\i}a At\'{o}%
mica, Av. Gral Paz 1499 (1650) San Mart\'{\i}n, Buenos Aires, Argentina}
\author{F. Parisi}
\affiliation{Departamento de F\'{\i}sica, Comisi\'{o}n Nacional de Energ\'{\i}a At\'{o}%
mica, Av. Gral Paz 1499 (1650) San Mart\'{\i}n, Buenos Aires, Argentina}
\affiliation{Escuela de Ciencia y Tecnolog\'{\i}a, UNSAM, Alem 3901, San Mart\'{\i}n,
Buenos Aires, Argentina}

\begin{abstract}
The occurrence at low temperatures of an ultrasharp field-induced transition
in phase separated manganites is analyzed. Experimental results show that
magnetization and specific heat step-like transitions below 5 K are
correlated with an abrupt change of the sample temperature, which happens at
a certain critical field. This temperature rise, a magnetocaloric effect, is
interpreted as produced by the released energy at the transition point, and
is the key to understand the existence of the abrupt field-induced
transition. A qualitative analysis of the results suggests the existence of
a critical growing rate of the ferromagnetic phase, beyond which an
avalanche effect is triggered.
\end{abstract}

\maketitle
\preprint{EPJB}

Mixed valent manganites show a great deal of fascinating properties, arising
from the strong interplay between spin, charge, orbital, and lattice degrees
of freedom \cite{Salamon}. The most intriguing one is the existence of a
phase separated state, the simultaneous coexistence of submicrometer
ferromagnetic (FM) metallic and charge ordered (CO) insulating regions \cite%
{Dagotto}. The phase separation scenario has its origin in the unusual
proximity of the free energies of these very distinct FM and CO states, and
in the fact that the competition between both phases is resolved in
mesoscopic length scales, giving rise to real space inhomogeneities in the
material.

Yet another surprising result more recently found in manganites is the
appearance of ultrasharp magnetization steps at low temperatures (below $%
\sim $ 5 K) in the isothermal magnetization $M$($H$) curves \cite{Hebert0,
Hebert1, Mahendiran, Hardy, Hebert2}. This effect, the field induced
transition of the entire compound from one phase to the other of the
coexisting states, is included in the category of metamagnetic transitions 
\cite{meta}. However, unlike the broad continuous transitions expected for
inhomogeneous granular systems, in this case it occurs in an extremely
narrow window of magnetic fields. These ultrasharp steps were observed in
both single crystals and polycrystalline samples, indicating that it is not
related to a particular micro-structure of the material.

The actual existence of a phase separated state was recognized as a key
parameter for the observation of these magnetization jumps \cite{Hebert2}.
The effect was first reported in manganites doped at the Mn site, and the
disorder in the spin lattice was thought to play a relevant role \cite%
{Hebert1}. However, a similar behavior was also found in Pr$_{0.6}$Ca$_{0.4}$%
MnO$_{3}$, and the qualitative interpretation of the phenomenon shifted to
the martensitic character of the phase separated state \cite{Hardy}.
Accommodation strains were shown to be relevant in the stabilization of
phase separation \cite{PodzorovMarten,Mathur}, but their role in the
magnetization steps is not clear, since it is expected that grain boundaries
would act as a sort of \textquotedblleft firewall\textquotedblright\ for the
movement of the domain walls, stopping the avalanche process. Additionally,
despite its intrinsic first-order character, the martensitic transformation
is spread over a large range of the external parameter driving the
transition, the magnetic field in the present case, in strong disagreement
with the abrupt character of the transition.

The aim of this investigations is to address a basic question concerning
this abrupt field-induced transition: why is this metamagnetic transition so
sharp, and what is actually causing it? We report the occurrence of
ultrasharp magnetization steps at low temperatures in a prototype phase
separated manganite, which are accompanied by discontinuities in the
magnetic field dependence of the specific heat. Concomitantly with these
facts, we found that the field-induced transition is accompanied by a large
increase in the temperature of the sample, by dozens of degrees. This
feature suggests a mechanism in which the abrupt first order transition in
the whole sample is triggered by the released heat in a microscopic phase
transformation. A low temperature heat controlled magnetization avalanche
was previously found in bulk disordered magnets \cite{UeharaStamp} due to
the heat released by the FM domain wall motion during the reversal of the
remnant magnetization. Also, local heating induced by non-uniform current
flow was proposed as the origin of the mesoscopic fluctuations between
coexisting phases observed in La$_{0.225}$Pr$_{0.40}$Ca$_{0.375}$MnO$_{3}$%
\cite{Podzorovfluctu}. We propose that in phase separated manganites the
interplay between the growth of the FM phase induced by the magnetic field
and the heat generated by this growth is the key to explain the avalanche
process leading to an ultrasharp field-induced transition in these
inhomogeneous strongly correlated systems.

The particular compound under study is a high quality polycrystalline sample
of La$_{0.225}$Pr$_{0.40}$Ca$_{0.375}$MnO$_{3}$, synthesized by the sol-gel
technique. It belongs to the well known family of compounds La$_{5/8-y}$Pr$%
_{y}$Ca$_{3/8}$MnO$_{3},$ whose tendency to form inhomogeneous structures in
the range 0.3$\leq y\leq $0.4 is extensively documented. \cite%
{Uehara,UeharaRate,Kim,Balagurov,Kiryukhin,nonvolatile} Scanning electron
micrographs revealed a homogeneous distribution of grain sizes, of the order
of 2 $\mu $m. An identification of the magnetic phases of the material can
be made through the temperature dependence of the magnetization, $M$($T$).
The results were obtained on an extraction magnetometer with a field $H$ = 1
T, and are shown in Fig. 1. As the temperature is lowered the sample changes
from a paramagnetic to a charge-ordered antiferromagnetic state at $T_{CO}$
= 220 K. Just below, a small kink at 190 K is a signature of the onset of
the formation of ferromagnetic clusters \cite{nonvolatile}. A more robust
ferromagnetic phase appears at $T_{C}$ = 70 K (90 K on warming), which
coexists with the majority CO state in an inhomogeneous phase separated
state \cite{Uehara}. In a temperature window extending from $T_{C}$ down to
a temperature $T_{b}$ $\simeq $ 20 K the magnetization shows considerable
relaxation effects, as shown in the inset of Fig. 1, signaling the growth of
the FM phase against the CO background. The temperature $T_{b}$ (which
depends on the applied field) can be identified as a blocking temperature;
relaxation below $T_{b}$ is strongly reduced. Additionally, the magnetic
state below $T_{b}$ is highly dependent on the sample magnetic field and
cooling history. If the sample is cooled without an applied field (ZFC) the
magnetization at 2 K shows a significant low value, that remains unchanged
while warming until $T_{b}$, above with it shows a continuous increase and
merges with the field cooling warming curve.

\begin{figure}[tbp]
\includegraphics[width=7cm,clip]{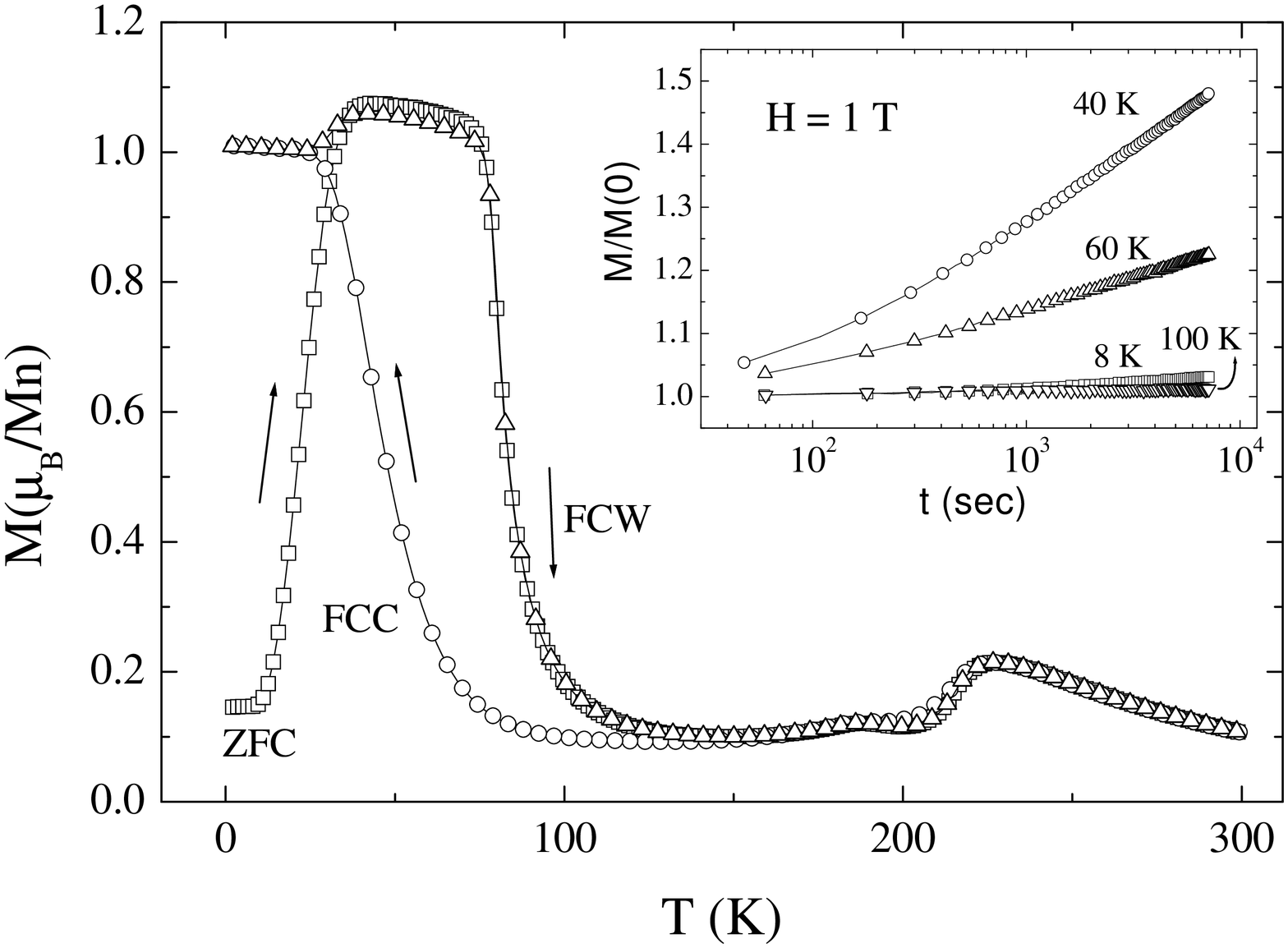}
\caption{ Temperature dependence of the   magnetization, measured with $H$ =
1 T with zero-field-cooling (ZFC), field-cooled-cooling (FCC), and
field-cooled-warming (FCW) procedures. The inset shows the time evolution of
the normalized magnetization after ZFC to $T$ = 8, 40, 60 and 100K.}
\label{Fig1}
\end{figure}

With the application of a large enough magnetic field the low temperature
(below $\sim $ 5 K) zero-field-cooled state is transformed into a FM phase
in an abrupt step-like metamagnetic transition. Figure 2(a) shows
magnetization measurements as a function of applied field, $M$($H$),
measured at $T$ = 2.5 K. At a certain critical field $H_{C}$ the entire
system changes to a nearly homogenous FM state, which remains stable even
after the field is removed. The width of the transition, determined by
repeating the measurements with lower field increments, is below 10 Oe.
Figure 2(b) shows specific heat data as a function of applied field, $C$($H$%
), measured by the relaxation method at the same base temperature, $T$ = 2.5
K. As can be readily noticed, a discontinuous transition occurs at
approximately the same magnetic field, indicating that a true thermodynamic
transition is taking place.

\begin{figure}[tbp]
\includegraphics[width=6cm,clip]{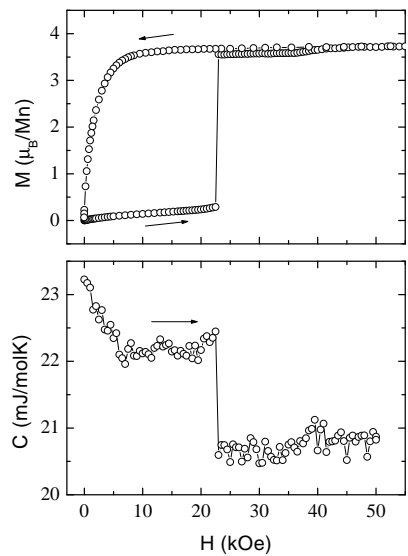}
\caption{(a) Field dependence of the magnetization and (b) specific heat,
measured at $T$ = 2.5 K. Both measurements show an abrupt change at the same
critical field $H_{C}$ $\simeq $ 2.2 T.}
\label{Fig2}
\end{figure}

Since the observed transition is first order, it is expected that the latent
heat involved should affect the thermodynamic state of the sample, for
instance, its temperature. In order to gain some insights on the magnitude
of the effect the following experiment was performed: with the sample placed
in a vacuum calorimeter with a weak thermal link to the temperature
controlled surroundings (kept at 2.5 K), the sample's temperature was
measured while the magnetic field was increased, with field increments
identical to the data of Fig.2. The obtained result, plotted in Fig. 3,
shows the occurrence of a sudden and huge increase of the sample's
temperature, greater than 25 K, at the same critical field of the
magnetization jump. Since the relaxation time for temperature stabilization
between the sample and the temperature controlled surroundings (of the order
of several seconds) is much larger than the internal time constant between
the sample and the sample holder (of the order of milliseconds), the
temperature rise measured is intrinsic to the sample. The abrupt increase of
the sample's temperature can be then doubtless ascribed to the heat
generated when the non-FM fraction of the material is converted to the FM
phase. The same experiment was repeated with samples of La$_{5/8-x}$Pr$_{x}$%
Ca$_{3/8}$MnO$_{3}$ with different Pr content, as well as in samples of the
series LaNdCaMnO. Whenever a magnetization jump occurs a sizable increase of
the sample's temperature was observed. It is also worth mentioned that in
the $M$($H$) and $C$($H$) data of figure 2 the sample's temperature is in
fact not strictly constant; there is also a sudden temperature rise at the
field of the step transition, followed by a quick relaxation to the base
temperature of system.

The process which starts with nearly the whole material in the non-FM phase
at $T_{0}$ = 2.5 K and ends with a nearly homogeneous FM state at $T_{f}$ $%
\approx $ 30 K, is conceptually related with the magnetocaloric (MC) effect 
\cite{Pecharsky}. 

\begin{figure}[tbp]
\includegraphics[width=6cm,clip]{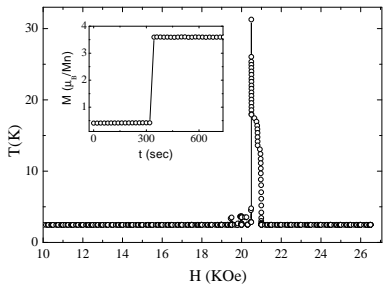}
\caption{ Field dependence of the sample's temperature showing an abrupt
warming from 2.5 to $\sim $ 30 K at $H_{C}$ $\simeq $ 2.2 T. The inset shows
a spontaneous magnetization jump, measured with a fixed magnetic field.}
\label{Fig3}
\end{figure}

The MC effect consists basically in a temperature change $\Delta T$ induced
by the application of a magnetic field, which, within the approximation of
reversible process, is related with the entropy change $\Delta S$ generated
by ordering the spin lattice. In our case, however, the approximation of
reversible adiabatic process is not valid, due to the strong irreversible
character of the field-induced transformation. Also, as the phase transition
from de CO to the FM phases involves large changes in the magnetic,
structural and electronic properties, which are strongly correlated, the
magnetic field affects all the degrees of freedom of the system. This fact
makes inapplicable some of the usual basic equations employed in the
description of the MC effect. A more realistic approach is to use the
conservation of the internal energy during the fast conversion process
(hypothesis of adiabaticity). Neglecting small changes of the sample volume,
we can make the identity $u_{CO}(T_{0})=u_{FM}(T_{f},H)$ , where $u$ is the
internal energy per volume unit. Replacing the whole (irreversible) process
by an isothermal plus an isobaric one, we can write: 
\begin{equation}
u_{CO}(T_{0})=u_{FM}(T_{0},H)+\int_{T_{0}}^{T_{f}}c_{p}dT
\end{equation}

where $c_{p}$ is the specific heat of the FM phase at constant pressure.
This yields an estimate for the released heat at the field induced
transition given by $\delta q_{rel}=u_{CO}(T_{0})-u_{FM}(T_{0},H)\simeq
\int_{T_{0}}^{T_{f}}c_{p}dT=48J/mol$. This estimated value was obtained from
specific heat measurements as a function of temperature performed at zero
magnetic field after the sample was transformed to the FM phase by
application of a field of 9 $T$.

The magnetization and specific heat results, shown in Fig. 2, are
macroscopic signatures of a phenomenon which must be understood at a
microscopic level. Below the temperature $T_{b}$ the sample gets into a
strongly blocked regime, in which the FM clusters can not grow against the
CO background (see inset of Fig 1). After zero- field-cooling, the sample
reaches the blocked state with a small, time independent, fraction $f$\ of
FM phase, which can be thought as distributed in isolated regions or
clusters surrounded by a CO matrix. The application of an external magnetic
field $H$ weakens this frozen-in state, inducing the increase of each
cluster of volume $v_{i}$ in an amount $\delta v_{i}$, which can depend on $%
v_{i}$, $T$, $H$ and time. The released heat yielded by this particular
process is $\delta Q_{rel}=q_{rel}\delta v_{i}$. Part of this energy is used
to locally increase the temperature of the FM volume $v_{i}$, a process that
can be considered as instantaneous, taking into account that the thermal
conductivity of the FM phase is much greater than that of the CO phase. The
remaining energy $\delta Q_{CO}$ is evacuated through the surrounding CO
region. This balance yields

\begin{equation}
q_{rel}\delta v_{i}=c_{p}v_{i}\delta T+\delta Q_{CO}
\end{equation}

Once a process involving a change of the local FM fraction happens, the
further evolution of the system is determined by the interplay between the
rates at which the system is generating heat, and the rate at which the CO
phase is releasing it. When the former is greater than the latter, a local
temperature rise within the FM region is obtained. If this temperature
reaches values beyond the blocking temperature corresponding to the applied
field $H$ the system becomes critical, in the sense that the adjacent CO
regions, which in turn will increase their local temperature too, become
highly unstable. These unstable CO regions are now easily transformed to the
FM state, releasing heat, and so on, inducing an avalanche-like chain
reaction. At the end, all regions which have ferromagnetism as its
equilibrium state at $T$=$T_{f}$ and field $H$ had been converted from CO to
FM. The equilibrium fractions of the coexisting phases at that $T$ and $H$
values then determine the size of the avalanche process.

Following equation (2), we can make an estimation of the critical value of
the volume change $\delta \nu _{i}^{crit}$\ which is needed to \textit{%
turn-on} the avalanche process. The first condition to be accomplished is
that the temperature of the local FM region increases beyond the blocking
temperature $T_{b}$($H$). Assuming that it occurs in a time scale $\tau $,
and that in this scale the heat transferred to the CO region is negligible,
we obtain

\begin{equation}
\delta v_{i}^{crit}/v_{i}=\frac{\int_{T_{0}}^{Tb}c_{p}dT}{q_{rel}}
\end{equation}

where $T_{b}(H)\simeq $ 8 K at $H_{c}$ = 2.2 T was estimated from ZFC
magnetization measurements. This calculation yields $\delta
v_{i}^{crit}/v_{i}\simeq 0.01$, i.e., almost one per cent increase of the
local FM volume is needed to initiate the abrupt transition. This condition
must be accompanied by another one, related with the time $\tau $ in which
the volume enlargement occurs. As mentioned above, this time has to be short
enough to avoid the heat be released through the surrounding CO region,
i.e., the condition $q_{rel}\delta v_{i}/\tau \gg a_{i}k\partial T/\partial
z $\ must hold, where $a_{i}$ is the area of the cluster surface, $k$ is the
thermal conductivity of the CO phase and $z$ is a local spatial coordinate.
A crude estimation of $\tau $ can be done assuming that, within the adjacent
CO region, the temperature decays from $T_{b}(H)$ to $T_{0}$ in a length $%
\delta v_{i}^{1/3}$, and taking into account that typical low temperature
values for $k$ could range between 0.1-1 W/(mK). These assumptions give, for
instance, $\tau \ll $10$^{-11}$-10$^{-12}$ s for clusters of volume $v_{i}$
= (50 nm)$^{3}$, and predicts a critical rate $\delta v_{i}^{crit}/\tau
\propto v_{i}^{1/3}$. Therefore, we estimate that thermal processes that
happen within a narrow time window, involving a one percent increase of the
local FM regions are needed to initiate the abrupt field-induced transition,
the critical rate scaling as the linear size of the FM cluster.

The occurrence of the step transition is then governed by the probability of
such an event, once the magnetic field has yielded the crossover between the
free energies of the coexisting phases. One way to modify the avalanche
probability is allowing the system to relax before reaching the critical
state. Allowing relaxation an increase of the FM fraction as a function of
the elapsed time occurs, and consequently the value of the $\delta
v_{i}^{crit}$\ needed to \textit{turn-on} the process should also increases.
This would be reflected on the dependence with the elapsed time of the
critical field $H_{C}$ at which the jump occurs. To verify this hypothesis
we have measured the time dependence of the magnetization during a field
sweep, $M$($H$,$t$) at 2.5 K, starting with $H$ = 1.9 T, and waiting a time $%
t_{w}$ at fixed $H$ before changing to the next field value. In Fig. 4 we
show the $M$ vs $t$ curves obtained for different values of $t_{w}$,
confirming the above presumption: for larger $t_{w}$ the magnetization jump
occurs at higher critical fields. As a remarkable result, we observed that
in most cases the step transition occurs spontaneously within the time
interval where the field was unchanged. This fact signs unambiguously that
the width of the step transition, beyond any experimental resolution, is
strictly zero. The inset of Fig. 3 shows a spontaneous (as opposed to
field-induced) magnetization jump, which happens at a fixed field and
temperature values. The occurrence of spontaneous magnetization jumps in
phase separated manganites was also reported by another group.\cite%
{spontaneous}

\begin{figure}[tbp]
\includegraphics[width=8.5cm,clip]{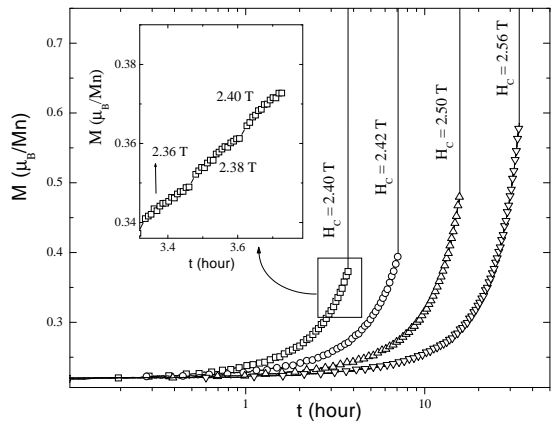}
\caption{Time dependence of the magnetization during a magnetic field sweep,
for different waiting times between consecutive field increments: $t_{w}$ =
7.5 (squares), 15 (circles), 30 (up triangles) and 60 min. (down triangles),
at $T$ = 2.5 K. The inset shows an enlarged portion of the region just
before the magnetization jump.}
\label{Fig4}
\end{figure}

The fact that the step transition can be reached spontaneously while the
external parameters ($H$ and $T$) are kept constant indicates that the
abrupt transition is truly connected with the probability of occurrence of
certain microscopic process, which within the above described scenario is a
particular enlargement of the FM phase. However, this process will initiate
the avalanche only when the local increment of the FM phase is large and
fast enough to yield the appropriate increase of the local temperature
through a magnetocaloric effect. Figure 4 and its inset clearly show that
not any enlargement process is able to trigger the step transition. The
relaxation effects displayed by the system before the occurrence of the
magnetization jump indicate that the system can in fact increase its FM
phase fraction without initiating the avalanche, i.e., there are FM regions
that starts to become unblocked for field values just below the critical
field, increasing their local volume by overcoming energy barriers. For
instance, the curve for $t_{w}$ = 60 min. (with $H_{C}$ = 2.56 T) shows a
sizable increase of the FM fraction before the occurrence of the
magnetization jump. From inspection of Fig. 4, it is likely that for larger
values of $t_{w}$ larger values of the FM fraction before the jump would be
obtained. The waiting time $t_{w}$ is a key parameter to determine the
energy barriers values for which the system is blocked. Eventually, for an
extremely large value of $t_{w}$ the whole system would behave as unblocked
and the $M$($H$) curve obtained in this hypothetical situation would display
a continuous metamagnetic transition behavior, without jumps. Therefore,
once the value of the minimal $H_{C}$ corresponding to the fastest
experiment is established, the limit temperature above which the step
transition no longer occurs is determined by the blocking temperature
corresponding to this field, $T_{b}$($H_{C}$). This suggests why the
magnetization jump occurs only below a very specific temperature,\cite%
{Mahendiran} above which the system overcomes the energy barriers without
turning on the avalanche process.

In conclusion, we have presented evidence that the low temperature abrupt
field-induced transition occurring in phase separated manganites is
intimately related with the sudden increase of the sample temperature at the
first order transition point, a feature which is crucial for the
understanding of the phenomenon. We proposed a simple model in which the
close interplay between the local increase of the FM phase and the heat
released in this microscopic transformation can turn-on the avalanche
leading to the observed step-like transition. Within this framework, the
entity which is propagated is heat, not magnetic domain walls, so the roles
of grain boundaries in ceramic samples or strains which exist between the
coexisting phases are less relevant. The observation of spontaneous
transitions gives additional support to that view, demonstrating that the
step transition is not only the result of a crossover between macroscopic
free energies induced by the magnetic field, but must be triggered by a
microscopic mechanism which initiates the avalanche process. Additionally,
we have established that a critical relative increment of a FM region or
cluster is needed for the system to reach the "chain reaction" state, i.e.,
larger initial FM factions require larger critical fields to \textit{turn-on}
the process, a feature previously observed. \cite{Mahendiran} Finally, it
must be emphasized that the basic condition for the occurrence of the abrupt
transition is that the system must reach the low temperature regime in a
strongly blocked state. At temperatures just a few degrees higher the abrupt
step-like transition no longer occurs, and it is replaced by a standard
continuous metamagnetic transition \cite{Mahendiran}. In summary, we propose
that some microscopic mechanism promotes locally a FM volume increase, which
yield a local temperature rise, and triggers the observed avalanche process.

This work was partially supported by CAPES, CNPq, FAPESP, CONICET, and
Fundaci\'{o}n Antorchas.

\section{Figure Caption}

Fig. 1. Temperature dependence of the magnetization, measured with $H$ = 1 T
with zero-field-cooling (ZFC), field-cooled-cooling (FCC), and
field-cooled-warming (FCW) procedures. The inset shows the time evolution of
the normalized magnetization after ZFC to $T$ = 8, 40, 60 and 100K.

Fig. 2. (a) Field dependence of the magnetization and (b) specific heat,
measured at $T$ = 2.5 K. Both measurements show an abrupt change at the same
critical field $H_{C}$ $\simeq $ 2.2 T.

Fig. 3. Field dependence of the sample's temperature showing an abrupt
warming from 2.5 to $\sim $ 30 K at $H_{C}$ $\simeq $ 2.2 T. The inset shows
a spontaneous magnetization jump, measured with a fixed magnetic field.

Fig. 4. Time dependence of the magnetization during a magnetic field sweep,
for different waiting times between consecutive field increments: $t_{w}$ =
7.5 (squares), 15 (circles), 30 (up triangles) and 60 min. (down triangles),
at $T$ = 2.5 K. The inset shows an enlarged portion of the region just
before the magnetization jump.

\end{document}